\documentclass[twocolumn]{aastex631}

\defcitealias{Rappaport1983}{RVJ}
\usepackage{MnSymbol}
\usepackage{newtxtext,newtxmath}
\usepackage[T1]{fontenc}
\usepackage{tikz}

\usepackage{wasysym}

\usepackage{xcolor}

\newcommand\orcid[1]{\href{http://orcid.org/#1}{\adjustbox{trim={-.15\width} {0\height} {-.15\width} {0\height},clip}{\includegraphics[height=12pt]{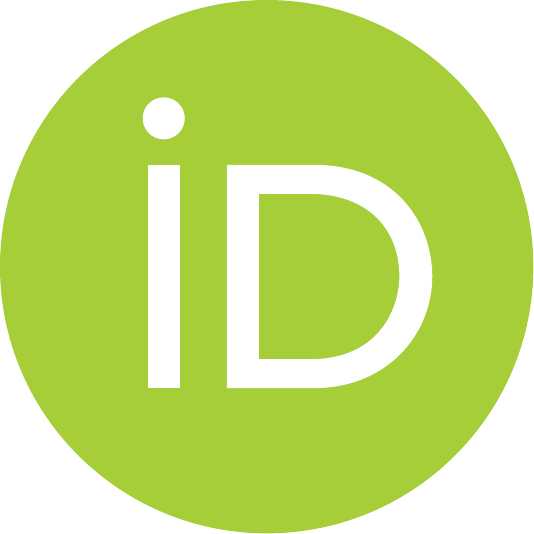}}}}

\usepackage{graphicx}	% Including figure files
\usepackage{amsmath}	% Advanced maths commands

\usepackage{listings}
\usepackage{color}
\definecolor{dkgreen}{rgb}{0,0.6,0}
\definecolor{gray}{rgb}{0.5,0.5,0.5}
\definecolor{mauve}{rgb}{0.58,0,0.82}
\definecolor{golden}{rgb}{0.86,0.65,0.01}
\begin{document}

% \title{Unlocking the Secrets of Low-Mass Star Formation: How Environment Shapes Stellar Births}
\title{The Variation of the Galaxy-Wide IMF for Low-Mass Stars: Modeling and Observational Insights}

\correspondingauthor{Zhiqiang Yan}
\email{yan@nju.edu.cn}

\author[0000-0001-7395-1198]{Zhiqiang Yan}
\affiliation{School of Astronomy and Space Science, Nanjing University, Nanjing 210093, People’s Republic of China}
\affiliation{Key Laboratory of Modern Astronomy and Astrophysics (Nanjing University), Ministry of Education, Nanjing 210093, People’s Republic of China}

\author[0000-0002-3651-5482]{Jiadong Li}
\affiliation{Max-Planck-Institut für Astronomie, Königstuhl 17, D-69117 Heidelberg, Germany}

\author[0000-0002-7301-3377]{Pavel Kroupa}
\affiliation{Helmholtz-Institut f{\"u}r Strahlen- und Kernphysik (HISKP), Universit{\"a}t Bonn, Nussallee 14–16, 53115 Bonn, Germany}
\affiliation{Charles University in Prague, Faculty of Mathematics and Physics, Astronomical Institute, V Hole{\v s}ovi{\v c}k{\'a}ch 2, 180 00 Praha 8, Czech Republic}

\author[0000-0002-1251-9905]{Tereza Jerabkova}
\affiliation{European Southern Observatory, Karl-Schwarzschild-Stra{\ss}e 2, 85748 Garching bei München, Germany}

\author[0000-0002-7440-1080]{Eda Gjergo}
\affiliation{School of Astronomy and Space Science, Nanjing University, Nanjing 210093, People’s Republic of China}
\affiliation{Key Laboratory of Modern Astronomy and Astrophysics (Nanjing University), Ministry of Education, Nanjing 210093, People’s Republic of China}

\author[0000-0002-7299-2876]{Zhi-Yu Zhang}
\affiliation{School of Astronomy and Space Science, Nanjing University, Nanjing 210093, People’s Republic of China}
\affiliation{Key Laboratory of Modern Astronomy and Astrophysics (Nanjing University), Ministry of Education, Nanjing 210093, People’s Republic of China}

%% automatically takes care of all commas and "and"s between authors' names.


\begin{abstract}
    The Stellar Initial Mass Function (IMF) characterizes the mass distribution of newly formed stars in various cosmic environments, serving as a fundamental assumption in astrophysical research. Recent findings challenge the prevalent notion of a universal and static IMF, proposing instead that the IMF's shape is contingent upon the star formation environment. 
    In this study, we analyze the galaxy-wide variation of the IMF for low-mass stars in both dwarf and massive galaxies with diverse observational methods. 
    Despite systematic discrepancies between different approaches, an IMF model with a metallicity-dependent slope for the low-mass stars aligns with the majority of observations, indicating a high degree of uniformity in the star formation processes across the universe. We also emphasize the need for a more comprehensive understanding of the variation of the low-mass IMF, considering measurement biases and factors beyond metallicity. 
\end{abstract}

%% You will be asked to select these concepts during the submission process


\section{Introduction}\label{sec: intro}

Massive stars play a pivotal role in shaping the dynamic and chemical evolution of galaxies, while low-mass stars serve as fossil records, preserving the history of galaxy formation. The crucial determinant of a star's life and legacy is its initial mass, making the distribution of initial masses a key factor in interpreting galaxy observations and conducting accurate galaxy simulations. However, measuring this distribution poses challenges, mainly due to the well-known degeneracy between the stellar initial mass function (IMF) and star formation history (SFH, \citealt{1979ApJS...41..513M,2009ApJ...706..599L}). The SFH affects the determination of the IMF but is often not tightly constrained for galaxies (and star clusters on a shorter timescale, \citealt{2011ApJ...727...64K,2023OJAp....6E..48G}).

Additionally, an ideal target for measuring the IMF of massive stars may not be suitable for studying the IMF of low-mass stars and vice versa, given the substantial luminosity difference between the two categories, reaching up to $10^8$ times. To address this challenge and obtain a comprehensive IMF across all stellar masses, researchers have employed various IMF measurement methods tailored to different stellar mass ranges in diverse systems (cf. the $\alpha$ plots in \citealt{2002Sci...295...82K} and \citealt{2017A&A...607A.126Y}). The underlying assumption guiding these efforts is that the IMF is a continuous function, following a universal formulation across different systems.

The challenge lies in unravelling the intricate connections between the star formation environment and the IMF \citep{2018PASA...35...39H,2020ARA&A..58..577S}. Variations of the IMF profoundly impact our interpretations of stellar systems, influencing estimates of their masses \citep{2023MNRAS.518.3494B,Haslbauer24}, ages \citep{2016MNRAS.456L.104M,2020ApJ...905...40S}, star formation rates \citep{2018A&A...620A..39J,2021arXiv211210788K}, dynamical evolution \citep{2021MNRAS.504.5778W,2022PASA...39....2N}, and chemical evolution \citep{2019A&A...629A..93Y,2021arXiv211208141Y,2021NatAs...5.1247M,2022MNRAS.516.3342W,2023A&A...669A.125T}.

Observations have consistently demonstrated that the IMF for massive stars is not universal but systematically varies with both the mass of a star cluster (\citealt{2012MNRAS.422.2246M,2022MNRAS.516.3342W,2023A&A...670A.151Y,2023ApJ...959...88D}) and the star formation rate (density) of a galaxy (\citealt{2009ApJ...706..599L,2011MNRAS.415.1647G,2012ApJ...747...72D,2017MNRAS.468.3071N,2018Natur.558..260Z,2021ApJ...923..120L,Guo24}). Encouragingly, the diverse IMFs observed on various spatial scales can be coherently explained by the integrated galactic IMF (IGIMF) theory (\citealt{2003ApJ...598.1076K,2013MNRAS.436.3309W,2017A&A...607A.126Y}). Furthermore, galaxy chemical evolution (GCE) calculations align with, and potentially favour, the scenario of a more top-heavy (top-light) IMF for massive (dwarf) galaxies (\citealt{1990ApJ...365..539M,1994A&A...288...57M,2017MNRAS.464.4866O,2019A&A...632A.110Y,2020A&A...637A..68Y,2021A&A...655A..19Y,2021NatAs...5.1247M}).

This work specifically delves into the IMF of low-mass stars ($<1~M_\odot$, where $M_\odot$ is the mass of the Sun), a challenging task due to their low luminosity. These stars dominate the total stellar mass in the present-day Universe, significantly impacting stellar mass estimations and shaping our comprehension of galaxy evolution (\citealt{2023arXiv231018464W}). The key question is whether the low-mass end of the IMF exhibits environmental dependencies akin to massive stars or if it manifests distinct behaviours. Additionally, we aim to explore the potential correlation between the IMFs of low and high-mass stars, investigating whether they vary independently in response to different physical parameters, as proposed by \citet{2018A&A...620A..39J}. It was pointed out there that for stars more massive than about $1\,M_\odot$, the stellar IMF depends on the metallicity and density of the embedded cluster. For low-mass stars, the stellar IMF appears to mostly depend on the metallicity \citep{2015ApJ...806L..31M,2018MNRAS.477.3954P,2019MNRAS.485.5256Z} while the relation between the shape of the galaxy-wide IMF of low-mass stars with other physical parameters such as the star-formation rate (SFR) may be only of secondary importance but is still unclear due to their long stellar lifetimes \citep{2019A&A...626A.124M}.

To test these proposals, we focus on the IMF of low-mass stars in old stellar populations, where massive stars do not dominate the total stellar luminosity. In such populations, a robust correlation between the shape of the IMF and stellar metallicity emerges (\citealt{2010Natur.468..940V,2013ApJ...771...29G,2015ApJ...806L..31M,2018MNRAS.477.3954P,2019MNRAS.485.5256Z,2020ARA&A..58..577S,2021ApJ...923...43V,2023Natur.613..460L} but see \citealt{2019MNRAS.489.4090L} and \citealt{2024arXiv240403939D}). Intriguingly, determinations of the IMF for low-mass stars within star clusters do not find a systematic variation of the IMF \citep{2021MNRAS.504.2557D,2023MNRAS.tmp..617B}. This uniformity in nearby star clusters may be attributed to relatively stable star-forming environments. It is also essential to consider that star clusters undergo mass segregation and dynamical evolution, making IMF estimations sensitive to factors such as initial cluster radii, gas expulsion, and assumptions regarding binary formation in star cluster models \citep{2007MNRAS.380.1589B,2012MNRAS.422.2246M,2021MNRAS.504.5778W,2023MNRAS.524.1422F}.

A similar situation arises in the study of ultra-compact dwarf galaxies (UCDs), where the formation process remains unclear \citep{2021MNRAS.502.5185M}. UCDs may have undergone a loss of outskirt stars during interactions with other galaxies in the past \citep{2003MNRAS.344..399B,2009MNRAS.394.1529D,2023Natur.623..296W}. To bridge the gap between low-mass IMF measurements in galaxies and star clusters, additional follow-up studies are imperative.

In this study, we explore the low-mass IMF in environments that are less affected by dynamical evolution, focusing on the galaxy-wide stellar IMF and field stars in the Milky Way (MW), and compare these observations with the IMF models with a metallicity-dependent variable shape proposed in \citet{2001MNRAS.322..231K,2002Sci...295...82K} and \citet{2020A&A...637A..68Y}. Specifically, we examine a linear correlation between the IMF slope and the stellar metal mass fraction, as outlined in Section~\ref{sec: models}.

The findings, detailed in Section~\ref{sec: observation}, reveal that the metallicity-dependent variable shape can effectively capture the primary variation trend of the shape of the IMF for low-mass stars in quiescent galaxies, even when these galaxies exhibit substantial differences in mass. The subsequent sections delve into theoretical considerations (Section~\ref{sec metallicity-dependent IMF}), a qualitative comparison between different models (Section~\ref{sec: Qualitative comparison between models}), discrepancies among various observations due to systematic biases of the IMF estimation method and additional physical parameters that affect the low-mass IMF (Section~\ref{sec: Discrepancies}), and additional studies estimating the IMF in star clusters and galaxies (Section~\ref{sec: Additional IMF studies}). A summary is provided in Section~\ref{sec: Conclusions}.

\section{The metallicity-dependent IMF}\label{sec: models}

\citet{2001MNRAS.322..231K,2002Sci...295...82K} first proposed an empirical variation law for the shape of the IMF of low-mass stars, grounded in observations of star clusters and field stars within the Solar neighbourhood and the Galactic bulge. This proposition suggests that the power-law indices of the IMF correlate with the logarithm of the stellar metal mass fraction, Z, following the relation:
% This proposition, tentatively supported by \citet{1999A&A...345..485P}, \citet{2000ApJ...534..870P}, and \citet{2001ApJ...546.1006B}, is expressed by the equation:
\begin{equation}\label{eq: Yan2020}
\alpha_{1,2}=\alpha_{1,2}^{\rm canonical}+c_1\cdot{\rm [Z]},
\end{equation}
where $\alpha_{1,2}$ denotes $\alpha_1$ and $\alpha_2$—the IMF power-law indices defined in equation~(\ref{eq: IMF}) for stars within the mass ranges of $0.08~M_\odot$ to $0.5~M_\odot$ and $0.5~M_\odot$ to $1~M_\odot$, respectively. Here, $\alpha_{1}^{\rm canonical}=1.3$ and $\alpha_{2}^{\rm canonical}=2.3$ \citep{2001MNRAS.322..231K}. The constant $c_1$ was estimated to be approximately $0.5$, suggesting that a relatively greater amount of low-mass stars are formed in gas clouds with a higher metallicity, denoted as ${\rm [Z]}\equiv{\rm log_{10}(Z/Z_\odot)}$ where ${\rm Z}_\odot=0.0142$ is the applied metal mass fraction of the Sun \citep{2009ARA&A..47..481A}. This observation-driven postulation aligns with the expectation that metal-rich clouds, characterized by more efficient cooling, would fragment into smaller clouds.

Considering that any influence of metal elements on the IMF should vanish as the metal mass fraction, Z, approaches zero, a new correlation between the IMF slope and Z was introduced in \citet{2020A&A...637A..68Y}:
\begin{equation}\label{eq: Yan2023}
\begin{split}
\alpha_{1,2}&=\alpha_{1,2}^{\rm canonical}+c_2\cdot({\rm Z-Z_{\rm MW}})\\
&=\alpha_{1,2}^{\rm canonical}+c_2\cdot(10^{[{\rm Z}]}-10^{\rm [Z]_{\rm MW}})\cdot{\rm Z_\odot}.
\end{split}
\end{equation}
Here, ${\rm Z_{\rm MW}}$ represents the mean metal mass fraction of field stars in the Solar neighbourhood, which served as the basis for canonical IMF measurements, that is, within about 20~pc \citep{1993MNRAS.262..545K,2010AJ....139.2679B,2024ApJS..271...55K}. Consequently, the $\alpha_{1,2}^{\rm canonical}$ value is automatically recovered when ${\rm [Z]}={\rm [Z]_{\rm MW}}\equiv{\rm log_{10}(Z_{\rm MW}/Z_\odot)}\approx-0.10\pm0.05$ (\citealt{2004A&A...418..989N,2021ApJS..253...45L,2023A&A...670A.107H} and \citealt{2023NatAs...7..951L}).
The constant $c_2=79.4$ is estimated by GCE modeling in \citet{2020A&A...637A..68Y,2021A&A...655A..19Y} and validated in \citet{2023A&A...669A.125T} (see equation~(\ref{eq: c2}) in Appendix~\ref{Appendix yan2020}).

According to the aforementioned postulations, with parameters summarized in Table~\ref{tab: models}, the IMF would exhibit a bottom-heavier trend, forming a relatively greater number of lower-mass stars as metallicity increases. This alignment with theoretical expectations is rooted in diverse perspectives on the origin of the IMF (\citealt{2013pss5.book..115K,2022MNRAS.509.1959S}).
\begin{table}
    \centering
    \caption{\label{tab: models}Parameters and equations of models for the variation of the IMF. $\Delta\alpha\equiv\alpha_{1}-\alpha_{1}^{\rm canonical}\equiv\alpha_{2}-\alpha_{2}^{\rm canonical}$ represents the deviation of the IMF power-law indices from the canonical IMF.}
    \begin{tabular}{ccc}
        \hline  \hline
        Model & Equation & $\Delta\alpha$\\
        \hline
        (i)     &  (\ref{eq: Yan2020}) & $0.5\cdot{\rm [Z]}$\\
        (ii)  &  (\ref{eq: Yan2023}) & $79.4\cdot({\rm Z-Z_{\rm MW}})$\\
        \hline
    \end{tabular}
\end{table}

It is important to note that the observed iron abundance, [Fe/H], of a dwarf galaxy is not directly comparable to the model calculation based on the total metallicity [Z]. However, considering a typical stellar-averaged $\alpha$-element enhancement, [$\alpha$/Fe], of less than 0.5~dex (\citealt{2017PASJ...69...76S}) and $\rm [Z]\approx[Fe/H]+[\alpha/Fe]$ \citep{2000AJ....120..165T}, the discrepancy is similar to the measurement uncertainty and would not significantly affect our comparison with the model for very metal-poor galaxies.

\section{Comparison with observations}\label{sec: observation}

\subsection{Comparison of the IMF slopes}\label{sec Compare the IMF slopes}

\subsubsection{A single power-law model for different mass ranges of stars}\label{sec: A single power-law fit}

The IMF is often characterized as a power law at its massive end, gradually flattening towards the low-mass range. When scrutinizing stars with masses approximately near $0.5~M_\odot$, where a mass boundary for the IMF to change its slope cannot be precisely determined, a common practice involves adopting a single power law to represent the IMF's shape (e.g. \citealt{2002NewA....7..395W,2013ApJ...763..110K,2018ApJ...855...20G,2023Natur.613..460L,2024arXiv240308850P}). To achieve this, we employ the Markov Chain Monte Carlo (MCMC) method. Specifically, we randomly sample one million stars from the assumed density distribution function, corresponding to any given IMF within a specific mass range. Subsequently, we apply a single power-law IMF fit to these stars.

The likelihood function for the power-law index of a galaxy-wide IMF is derived by assuming that each stellar mass is independently distributed, with the probability of obtaining a specific mass determined by the value of the galaxy-wide IMF (defined as the number of stars per stellar mass interval) at that particular mass:
\begin{equation}\label{eq: IMF}
    \xi(m) = \frac{{\rm d}N}{{\rm d}m} = A m^{-\alpha}.
\end{equation}
The likelihood function $L$ is expressed as the product of the probabilities for each individual stellar mass $m_i$:
\begin{equation}\label{eq: likelihood}
    L = \prod^N_{i=1} p (m_i \mid \theta) = \prod^N_{i=1} \xi (m_i, \theta).
\end{equation}
Here, $\theta$ represents the parameters, including the normalization parameter $A$ and the power-law index $\alpha$. The log$_{10}$-likelihood can then be expressed as:
\begin{equation}\label{eq:loglikelihood}
    \log L = \log_{10} A - \alpha \sum_{i=1}^N \log_{10} m_i.
\end{equation}
The total number of stars $N$ can be determined by integrating the galaxy-wide IMF over the stellar mass range $m_{\rm low}$ to $m_{\rm up}$:
\begin{equation}\label{eq: N}
    N = \int_{m_{\rm low}}^{m_{\rm up}} A m^{-\alpha} \mathrm{d}m = \frac{m_{\rm up}^{1-\alpha} - m_{\rm low}^{1-\alpha}}{1 - \alpha} A,
\end{equation}
This allows us to express the parameter $A$ as a function of $N$, $m_{\rm low}$,  $m_{\rm up}$, and $\alpha$ in the log$_{10}$-likelihood equation~(\ref{eq:loglikelihood}).

For our MCMC sampling in this study, we utilized the {\tt\string NumPyro} package \citep{phan2019composable}. The parameters $A$ and $\alpha$ were assigned flat priors, and the sampling process involved 200 warm-up steps followed by 800 fitting steps. Subsequently, we obtained the mean and standard deviation of the resulting $\alpha$ values. In cases where the standard deviation of $\alpha$ exceeded 0.01, additional iterations with a larger number of stars were conducted to refine the calculation and decrease the standard deviation.

\subsubsection{Observed IMF slopes in dwarf galaxies}

\begin{figure}\centering
    \includegraphics*[width=\columnwidth]{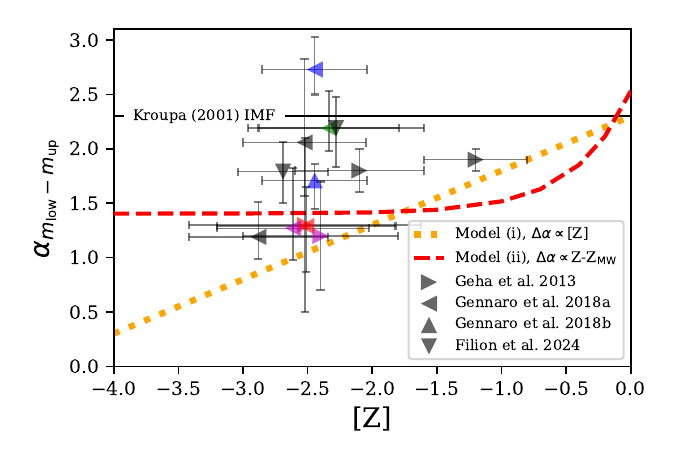}
    \caption{\label{fig: dwarfs}The power-law indices of the IMF, $\alpha$, for stars between specific stellar mass ranges, $m_{\rm low}$ and $m_{\rm up}$, for nearby UFDs and the SMC (data listed in Table~\ref{tab: data}), in comparison with single power-law slopes for stars above $0.5~M_\odot$ expected by Model (i) shown as the orange dotted line and Model (ii) shown as the red dashed curve (models summarized in Table~\ref{tab: models}).
    The horizontal solid line denotes the value of the canonical IMF given by \citet{2001MNRAS.322..231K}. Model (i) and Model (ii) recover the canonical IMF at [Z]=0 and [Z]=-0.1, respectively (see Section~\ref{sec: models}).
    For repeated studies on the same galaxy, the triangles coloured blue, red, and magenta, represent the Coma Berenices, Leo IV, and Hercules UFD, respectively. The green triangle represents the Bo\"otes I UFD.
    }
\end{figure}

\citet{2013ApJ...771...29G}, \citet{2018ApJ...855...20G,2018ApJ...863...38G}, and \citet{2024arXiv240411571F} conducted measurements of the galaxy-wide IMF for low-mass stars in slightly different mass ranges in nearby ultra-faint dwarf galaxies (UFDs). They employed a star count on the colour-magnitude diagram (CMD) below the main-sequence turnoff.
We compile these IMF estimations in Table~\ref{tab: data}. 
For \citet{2018ApJ...855...20G,2018ApJ...863...38G} that only provide IMFs for binary stellar systems, the IMFs for single stars are calculated with a uniform mass ratio distribution following their assumption. The results are plotted in Figure~\ref{fig: dwarfs} in comparison with different models of the IMF.
We also calculate the IMF solutions that deviate from the canonical IMF slope by $\Delta\alpha\equiv\alpha_{1}-\alpha_{1}^{\rm canonical}\equiv\alpha_{2}-\alpha_{2}^{\rm canonical}$ (parameters defined in Equations~(\ref{eq: Yan2020}) and (\ref{eq: Yan2023})) and have a best-fit single power-law IMF slope (Section~\ref{sec: A single power-law fit}) the same as the observed slope in the observed stellar mass range.
Overall, the Small Magellanic Cloud (SMC) and nearby UFDs suggest negative $\Delta\alpha$. That is, a shallower IMF for low-mass stars, in line with a metallicity variation as suggested here.
\begin{table*}
    \centering
    \caption{\label{tab: data}The IMF slopes from literature plotted in Figure~\ref{fig: dwarfs}.}
    \begin{tabular}{cccccc}
        \hline  \hline
        Identifier & [Z] & Mass range [$M_\odot$] & $\alpha_{m_{\rm low}-m_{\rm up}}$ & $\Delta\alpha$ &  Reference\\
        \hline
        Hercules & $-2.4\pm0.6$ & 0.52 to 0.76 & $1.2\pm0.5$ & $-1.1\pm0.5$ & Geha et al. (2013)\\
        Leo IV & $-2.52\pm0.9$ & 0.54 to 0.77 & $1.3\pm0.8$ & $-1.0\pm0.8$ & Geha et al. (2013)\\
        Ursa Minor & $-2.1\pm0.5$ & 0.40 to 0.80 & $1.8\pm0.2$ & $-0.2\pm0.2$ & Geha et al. (2013)\\
        SMC & $-1.2\pm0.4$ & 0.37 to 0.93 & $1.9\pm0.1$ & $-0.1\pm0.1$ & Geha et al. (2013)\\
        
        Bo\"o I & $-2.34\pm0.54$ & 0.41 to 0.77 & $2.26\pm0.5$ & $0.2\pm0.6$ & Gennaro et al. (2018a)\\
        Com Ber & $-2.45\pm0.40$ & 0.35 to 0.69 & $2.70\pm0.6$ & $1.0\pm0.6$ & Gennaro et al. (2018a)\\
        UMa I & $-2.53\pm0.47$ & 0.52 to 0.79 & $2.15\pm1.0$ & $-0.1\pm1.0$ & Gennaro et al. (2018a)\\
        CVn II & $-2.88\pm0.54$ & 0.43 to 0.81 & $1.23\pm0.3$ & $-0.9\pm0.4$ & Gennaro et al. (2018a)\\
        Hercules & $-2.61\pm0.59$ & 0.50 to 0.80 & $1.31\pm0.7$ & $-1.0\pm0.8$ & Gennaro et al. (2018)\\
        Leo IV & $-2.51\pm0.69$ & 0.51 to 0.80 & $1.26\pm0.7$ & $-1.0\pm0.8$ & Gennaro et al. (2018a)\\

        Com Ber & $-2.45\pm0.40$ & 0.25 to 0.75 & $1.71\pm0.5$ & $0.2\pm0.5$ & Gennaro et al. (2018b)\\
        
        Ret II & $-2.69\pm0.35$ & - & $1.79\pm0.3$ & $0.2\pm0.3$ & Filion et al. (2024)\\
        UMa II & $-2.28\pm0.68$ & - & $2.19\pm0.4$ & $0.6\pm0.4$ & Filion et al. (2024)\\
        \hline
    \end{tabular}
    \tablerefs{\citet{2013ApJ...771...29G,2018ApJ...855...20G,2018ApJ...863...38G,2024arXiv240411571F}.}
    \tablecomments{The power-law indices for the IMF of single stars, $\alpha$, assuming a single-power-law IMF, are estimated with observed individual stars (or unresolved stellar systems) with a mass between $m_{\rm low}$ and $m_{\rm up}$. The completeness limit for the apparent magnitude is provided in \citet{2024arXiv240411571F}, roughly corresponding to a stellar mass range of 0.2 to $1~M_\odot$. 
    $\Delta\alpha$ provides a uniform measure to compare the IMF slopes. A single-power-law fit (Section~\ref{sec: A single power-law fit}) for stars between $m_{\rm low}$ and $m_{\rm up}$ with a mass distribution described by equation~(\ref{eq: Yan2020}) or (\ref{eq: Yan2023}) that deviates from the canonical \citet{2001MNRAS.322..231K} IMF by $\Delta\alpha\equiv\alpha_{1}-\alpha_{1}^{\rm canonical}\equiv\alpha_{2}-\alpha_{2}^{\rm canonical}$ would result in the observed slope $\alpha_{m_{\rm low}-m_{\rm up}}$. A negative $\Delta\alpha$ value reflects a bottom-lighter IMF than the canonical IMF.}
\end{table*}

The large scatter of measured IMF slopes may have several origins.
First of all, the IMF has a stellar mass-dependent slope (as illustrated in figure 5 of \citealt{2002Sci...295...82K} where stars more massive than about $0.5~M_\odot$ exhibit steeper IMF slopes than lower-mass stars) while the observed stellar samples in different UFDs are in different mass ranges ($m_{\rm low}$ to $m_{\rm up}$ listed in Table~\ref{tab: data}). Therefore, the $\alpha_{m_{\rm low}-m_{\rm up}}$ values of these galaxies are expected to be different by about 0.2 even if they have identical IMFs.
In addition, the disagreement between the UFDs and Model~(ii) may suggest that the stellar binary fractions of these galaxies are overestimated or additional physical factors other than metallicity are affecting the shape of the IMF as discussed in Section~\ref{sec: Discrepancies}.

\subsubsection{Observed IMF slopes in massive early-type galaxies}

\begin{figure}\centering
    \includegraphics*[width=\columnwidth]{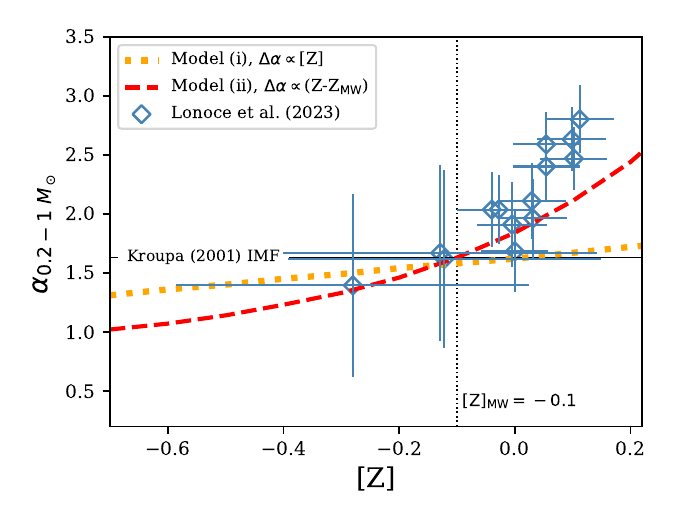}
    \caption{\label{fig: Lonoce23}A comparison for stars within the mass range of $0.2~M_\odot$ to $1~M_\odot$, contrasting model predictions with observations of an old ETG, NGC3309 \citep{2023ApJ...948...65L}. 
    The models and the canonical Kroupa IMF are the same as in Figure~\ref{fig: dwarfs} but for a different stellar mass range.
    The vertical dotted line indicates the mean stellar metallicity in the Solar neighbourhood, where the canonical IMF slope is recovered.
    }
\end{figure}

The low-mass IMF slope of massive early-type galaxies (ETGs) can be determined through stellar population synthesis (SPS). 
As an illustrative case, we consider the IMF measurement of NGC3309 by \citet{2023ApJ...948...65L} and compare it with different models in Figure~\ref{fig: Lonoce23}. We note that galaxy NGC3311 from the same study is not included here due to its non-negligible star formation activity. The ETG NGC3309 has an average stellar age of approximately 12~Gyr. Considering the stellar lifetime and the low luminosity of low-mass stars, we assume that the determination of the galaxy-wide IMF based on SPS is primarily sensitive to the stellar mass range from $0.2~M_\odot$ to $1~M_\odot$, as suggested by \citet{2019A&A...626A.124M}. 

In Figure~\ref{fig: Lonoce23}, the results from \citet{2023ApJ...948...65L} are compared with a single power-law fit (Section~\ref{sec: A single power-law fit}) of the IMF models in the mass range $0.2~M_\odot$ to $1~M_\odot$. The standard deviation of the best-fit value of the single power-law index, $\alpha$, is smaller than 0.01. Model (ii), with $\Delta\alpha\propto {\rm Z}$, agree best with the observed bottom-heavy IMFs of metal-rich stellar populations.

\subsubsection{Summary of observed low-mass IMF slopes}

The above comparison demonstrates that Model (ii) best accommodates observational constraints across various galactic metallicities. Notably, the observations are derived from diverse methods, including direct star counts \citep{2013ApJ...771...29G}, GCE \citep{2020A&A...637A..68Y}, and SPS \citep{2023ApJ...948...65L}.

However, different IMF studies observe stars within distinct stellar mass ranges. A fair comparison between their IMFs is hindered by the inherent variation in IMF slopes with stellar mass \citep[cf.,][their section 3.3.1]{2021A&A...654A..59M}. Thus, different galaxies in Figure~\ref{fig: dwarfs} and \ref{fig: Lonoce23}, presenting the IMF slopes for stars within significantly different mass ranges, cannot be directly compared in the same plot. This problem can be circumvented by extrapolating the measured IMFs and comparing homogeneously the mass ratios of stars in given mass ranges.

\subsection{Comparison of the stellar mass ratios}\label{sec Stellar mass ratio}

Substantial IMF studies report a correlation between the slope of the low-mass IMF and stellar metallicity. However, these studies often assume different IMF formulations, posing challenges for a direct comparison. For instance, \citet{2015ApJ...806L..31M,2019A&A...626A.124M,2021A&A...654A..59M} and \citet{2018MNRAS.478.4084S} adopt the ``bimodal low-mass tapered'' IMF introduced by \citet{1996ApJS..106..307V}, incorporating a variable IMF slope parameter for stars above $0.6~M_\odot$. The variation in the high-mass IMF indirectly influences the IMF for stars in the range of 0.2 to $0.6~M_\odot$ through a spline connection between the high and low-mass IMFs in this formulation. Conversely, some studies employ a two-part power-law IMF with a variable slope for stars less massive than $0.5~M_\odot$ \citep{2018MNRAS.477.3954P} or $1~M_\odot$ \citep{2023ApJ...948...65L}. 
These different IMF formulations in studies of potential systematic variation of the IMF are necessary assumptions because the detailed shape of the IMF does not have a unique solution.

\begin{figure}\centering
    \includegraphics*[width=\columnwidth]{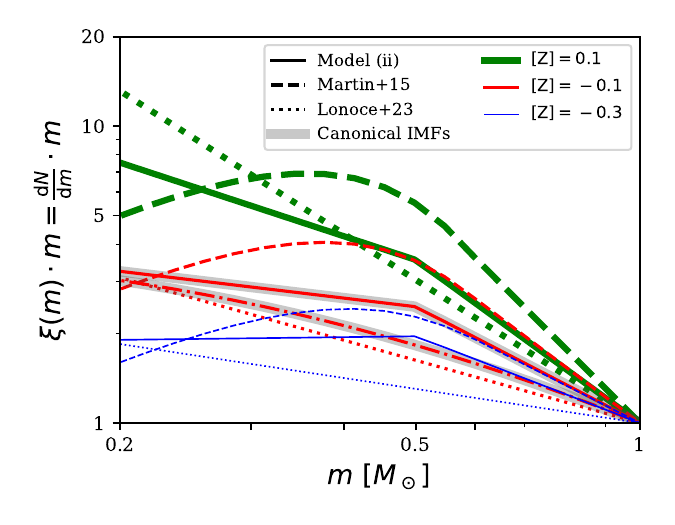}
    \caption{\label{fig: IMFs}Comparison of Model (ii) with observational IMF determinations based on different IMF formulations as priors. The stellar metallicity is indicated by the line's colour and thickness (see inset). Solid lines depict Model (ii). Observations for ETGs with different metallicities from \citet{2015ApJ...806L..31M}, assuming the ``bimodal low-mass tapered'' IMF, are shown by the dashed curves. Observations of NGC3309 from \citet[their figure 11]{2023ApJ...948...65L}, assuming a single power-law IMF below $1~M_\odot$, are shown by the dotted lines. Canonical power-law and lognormal IMFs from \citet[red solid line]{2001MNRAS.322..231K} and \citet[red dash-dotted curve]{2003PASP..115..763C} are highlighted by a grey background. We note that Model (ii) recovers the canonical Kroupa IMF at the metallicity of the Solar neighbourhood, $[{\rm Z}]=-0.1$. All IMFs are normalized to have the same value at $m=1~M_\odot$.
    }
\end{figure}

% To transform these diverse IMF estimations into the parameter $\xi_{\rm MR}$, we follow the IMF assumptions of individual studies. 
We demonstrate several examples of power-law IMFs and ``bimodal low-mass tapered'' IMFs for stars with different metallicities (${\rm [Z]}=0.1$, $-0.1$, and $-0.3$) in Figure~\ref{fig: IMFs}. The ``bimodal low-mass tapered'' IMFs shown by the dashed curves follow the metallicity-dependent IMF slope given by \citet[their equation 2]{2015ApJ...806L..31M}. The single-power-law IMF shown by the dotted lines have slopes taken from observations of NGC3309 from the right panel of figure 11 of \citet{2023ApJ...948...65L}. Model (ii) is shown by the solid lines. We note that any variable model of the IMF should be consistent with the shape of the IMF deduced from the Solar-neighbourhood stars (grey backgrounds in Figure~\ref{fig: IMFs}).

To facilitate a meaningful comparison between the measurements assuming different formulations of the IMF, we calculate the parameter $\xi_{\rm MR}$, representing the mass ratio of stars within the ranges of $0.2~M_\odot$ to $0.5~M_\odot$ and $0.2~M_\odot$ to $1~M_\odot$, as defined in \citet{2019A&A...626A.124M}. Namely, for a stellar IMF defined as $\xi(m)$, $\xi_{\rm MR}\equiv\int_{0.2~M_\odot}^{0.5~M_\odot} \xi(m)\cdot m~{\rm d}m \div \int_{0.2~M_\odot}^{1~M_\odot} \xi(m)\cdot m~{\rm d}m$.

% Biases in estimating the $\xi_{\rm MR}$ value are introduced not only by the different assumed IMF formulations shown in Figure~\ref{fig: IMFs} but also by the mass ranges of the observed stars. 
For star-counting studies in a mass range ($m_{\rm low}$ to $m_{\rm up}$) narrower than $0.2~M_\odot$ to $1~M_\odot$ \citep{2013ApJ...771...29G,2018ApJ...855...20G,2018ApJ...863...38G,2024arXiv240411571F}, we assume the broken power-law IMF solution with the best-fit single power-law IMF slope the same as the observed value in the observed stellar mass range ($\alpha_{m_{\rm low}-m_{\rm up}}$ in Table~\ref{tab: data}). Then, we calculate the $\xi_{\rm MR}$ value for this best-fit IMF.
For \citet{2023Natur.613..460L} on the IMF of MW field stars, we do not extrapolate the IMF slope but assume a tripartite IMF with power-law indices $-1.3$, the observational best-fit slope, and $-2.3$ for stars below $m_{\rm low}$, $m_{\rm low}$ to $m_{\rm up}$, and above $m_{\rm up}$, respectively, following \citet{2001MNRAS.322..231K}. These assumptions are the simplest to compare different IMF observations and their systematic variation.

\begin{figure*}\centering
    \includegraphics[width=\textwidth]{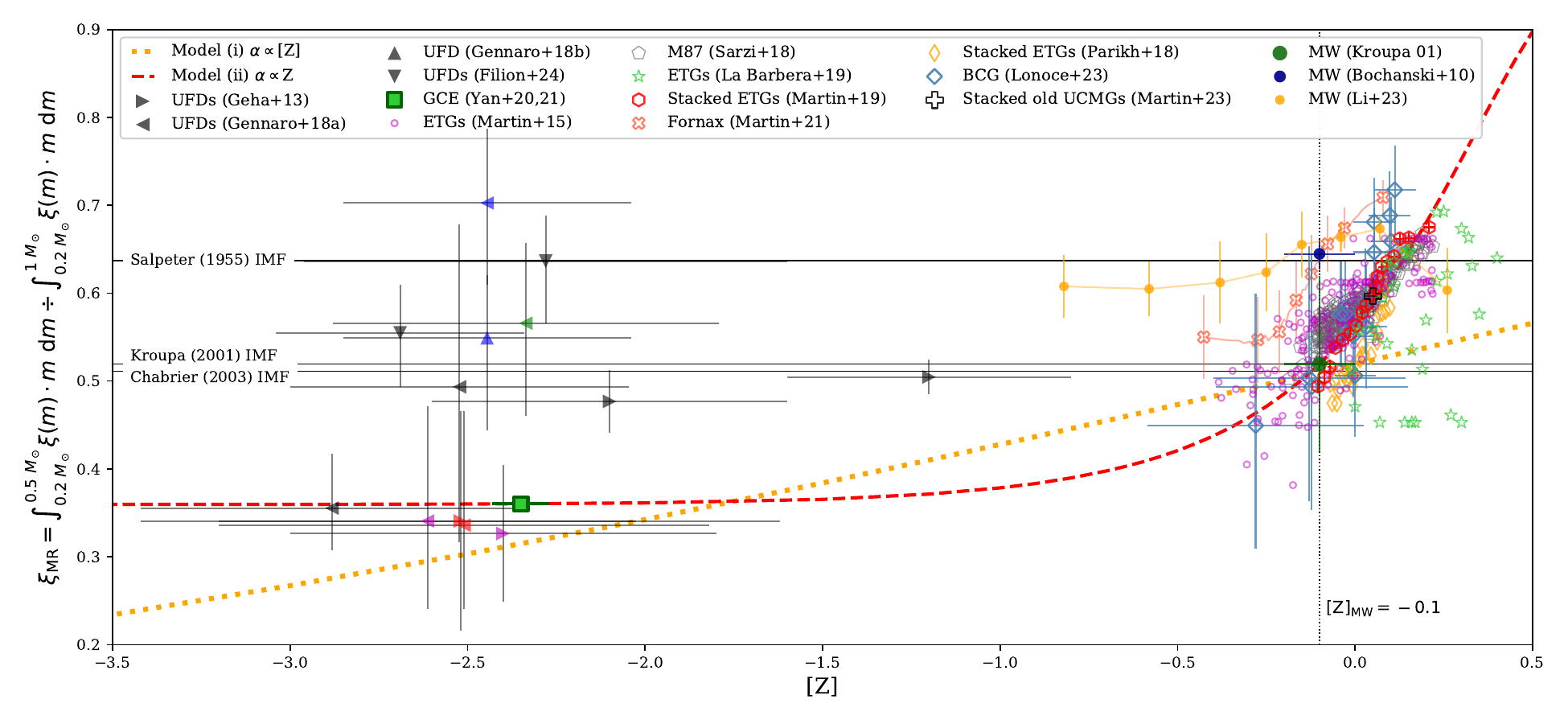}
    \caption{\label{fig: low_mass_IMF}The observed mass ratio in stars between $0.2~M_\odot$ to $0.5~M_\odot$ and $0.2~M_\odot$ to $1~M_\odot$, $\xi_{\rm MR}$, in comparison with the expected value of our models (Table~\ref{tab: models}) as a function of stellar metallicity.
    Independent IMF determination methods are considered, including direct star counts \citep{2001MNRAS.322..231K,2010AJ....139.2679B,2013ApJ...771...29G,2018ApJ...855...20G,2018ApJ...863...38G,2023Natur.613..460L}, SPS \citep{2015ApJ...806L..31M,2019A&A...626A.124M,2021A&A...654A..59M,2023MNRAS.521.1408M,2018MNRAS.477.3954P,2018MNRAS.478.4084S,2019MNRAS.489.4090L,2023ApJ...948...65L} of resolved or stacked quiescent galaxies, and GCE \citep{2020A&A...637A..68Y,2021A&A...655A..19Y}, as summarized in Table~\ref{tab: data2}. These studies assume different formulations of IMF (Figure~\ref{fig: IMFs}) which leads to systematically different $\xi_{\rm MR}$ values (see discussions in Section~\ref{sec: Discrepancies}). The error bars for data points from \citet{2015ApJ...806L..31M,2018MNRAS.478.4084S,2018MNRAS.477.3954P,2019MNRAS.489.4090L} are provided in the original papers but are omitted here for a cleaner illustration. The GCE study from \citet{2020A&A...637A..68Y} did not provide an IMF estimation uncertainty as is explained in Appendix~\ref{Appendix yan2020}. The colours of the triangles stand for different UFDs, the same as in Figure~\ref{fig: dwarfs}. The horizontal lines mark $\xi_{\rm MR}=0.637$, $0.5194$, $0.5114$ for \citet[Eq. 5 therein]{1955ApJ...121..161S}, \citet[Eq. 2 therein]{2001MNRAS.322..231K}, and \citet[Eq. 17 therein]{2003PASP..115..763C} IMFs, respectively.
    }
\end{figure*}

With $\xi_{\rm MR}$, we extend our comparison in Section~\ref{sec Compare the IMF slopes} to incorporate additional observations from \citet{2001MNRAS.322..231K,2010AJ....139.2679B,2015ApJ...806L..31M,2019A&A...626A.124M,2021A&A...654A..59M,2023MNRAS.521.1408M,2018MNRAS.477.3954P,2018MNRAS.478.4084S}, and \citet[excluding the relatively younger galaxy XSG6]{2019MNRAS.489.4090L}.
% , and brightest cluster galaxies (BCGs) from \citet[excluding the UCDs and GCs that are affected by dynamical evolution]{2023MNRAS.526.4004C} 
The observations are summarized in Table~\ref{tab: data2} and visualized in Figure~\ref{fig: low_mass_IMF}.
Generally, stellar populations with metallicity similar to the MW, denoted by the vertical dotted line, align with the canonical single-star IMFs from \citet[Eq.~5 therein]{1955ApJ...121..161S}, \citet[Eq.~2 therein]{2001MNRAS.322..231K}, and \citet[Eq.~17 therein]{2003PASP..115..763C} (as indicated by horizontal lines). 
% The $\xi_{\rm MR}$ uncertainty of the \citet{2001MNRAS.321..699K} IMF is calculated assuming a power-law indices variation of $\pm0.3$ according to \citet{2013pss5.book..115K}.
A metallicity--IMF correlation is evident across the observational studies, with a steeper IMF slope for metal-rich low-mass stars.
Notably, the stacked spectra of old ETGs and old ultra-compact massive galaxies (UCMGs) analyzed by \citet{2019A&A...626A.124M} and \citet{2023MNRAS.521.1408M}, respectively, exhibit remarkably high signal-to-noise ratios and small uncertainties, closely tracking Model (ii). In contrast, uncertainties in the other studies for a single galaxy or galaxy regions are generally much larger. 

\begin{table*}
    \caption{\label{tab: data2}Data compiled for Figure~\ref{fig: low_mass_IMF}.
    }
    \centering
    \begin{tabular}{ccccccccc}
        \hline  \hline
        Target & Identifier & [Z] & Method & Mass range [$M_\odot$] & IMF & $\xi_{\rm MR}$ & Reference\\
        \hline
        UFD & Hercules & $-2.4\pm0.6$ & CS & 0.52 to 0.76 & SPL & $0.33\pm0.08$ & Geha et al. (2013)\\
        UFD & Leo IV & $-2.52\pm0.9$ & CS & 0.54 to 0.77 & SPL & $0.34\pm0.13$ & Geha et al. (2013) \\
        UFD & Ursa Minor & $-2.1\pm0.5$ & CS & 0.4 to 0.8 & SPL & $0.48\pm0.04$ & Geha et al. (2013) \\
        Dwarf & SMC & $-1.2\pm0.4$ & CS & 0.37 to 0.93 & SPL & $0.50\pm0.02$ & Geha et al. (2013) \\

        UFD & Bo\"o I & $-2.34\pm0.54$ & CS & 0.41 to 0.77 & SPL & $0.57\pm0.10$ & Gennaro et al. (2018a)\\
        UFD & Com Ber & $-2.45\pm0.40$ & CS & 0.35 to 0.69 & SPL & $0.70\pm0.09$ & Gennaro et al. (2018a) \\
        UFD & UMa I & $-2.53\pm0.47$ & CS & 0.52 to 0.79 & SPL & $0.49\pm0.18$ & Gennaro et al. (2018a)\\
        UFD & CVn II & $-2.88\pm0.54$ & CS & 0.43 to 0.81 & SPL & $0.36\pm0.06$ & Gennaro et al. (2018a) \\
        UFD & Hercules & $-2.61\pm0.59$ & CS & 0.50 to 0.80 & SPL & $0.34\pm0.12$ & Gennaro et al. (2018a) \\
        UFD & Leo IV & $-2.51\pm0.69$ & CS & 0.51 to 0.80 & SPL & $0.34\pm0.12$ & Gennaro et al. (2018a) \\
        
        UFD & Com Ber & $-2.45\pm0.40$ & CS & 0.25 to 0.75 & SPL & $0.55\pm0.10$ & Gennaro et al. (2018b) \\
        
        UFD & Ret II & $-2.69\pm0.35$ & CS & - & SPL & $0.55\pm0.06$ & Filion et al. (2024)\\
        UFD & UMa II & $-2.28\pm0.68$ & CS & - & SPL & $0.64\pm0.06$ & Filion et al. (2024) \\
        
        UFD & Bo\"o I & $-2.35\pm0.08$ & GCE & - & BPL & $0.36$ & Yan et al. (2020, 2021) \\

        ETG & ETGs & -0.38 to 0.22 & SPS & - & BT & 0.38 to 0.66 & Mart\'in-Navarro et al. (2015) \\
        ETG & M87\tablenotemark{$^a$} & -0.19 to 0.22 & SPS & - & BT & 0.50 to 0.67 & Sarzi et al. (2018) \\
        ETG & ETGs\tablenotemark{$^a$} & 0 to 0.4 & SPS & - & BT & 0.45 to 0.69 & La Barbera et al. (2019) \\
        ETG & ETGs\tablenotemark{$^b$} & -0.1 to 0.2 & SPS & - & BT & 0.49 to 0.68 & Mart\'in-Navarro et al. (2019) \\
        QGs & Fornax\tablenotemark{$^a$} & -0.45 to 0.1 & SPS & - & BT & 0.5 to 0.72 & Mart\'in-Navarro et al. (2021) \\
        ETG & ETGs\tablenotemark{$^{ab}$} & -0.07 to 0.1 & SPS & - & BPL & 0.47 to 0.62 & Parikh et al. (2018) \\
        BCG & NGC3309\tablenotemark{$^a$} & -0.28 to 0.11 & SPS & - & SPL & 0.45 to 0.72 & Lonoce et al. (2023) \\
        UCMG & Relics\tablenotemark{$^b$} & $0.05\pm0.02$ & SPS & - & BT & $0.60\pm0.02$ & Mart\'in-Navarro et al. (2023) \\
        MW & Field stars & $-0.1\pm0.1$ & CS & 0.1 to 1 & BPL & $0.52\pm0.10$ & Kroupa (2001) \\
        MW & Field stars & $-0.1\pm0.1$ & CS & 0.1 to 0.79 & BPL & $0.645\pm0.018$ & Bochanski et al. (2010) \\
        MW & Field stars & -0.82 to 0.26 & CS & 0.3 to 0.7 & SPL & 2.29 to 2.77 & Li et al. (2023) \\
        \hline
    \end{tabular}
    \tablerefs{\citet{2013ApJ...771...29G,2018ApJ...855...20G,2018ApJ...863...38G,2024arXiv240411571F,2020A&A...637A..68Y,2021A&A...655A..19Y,2018MNRAS.478.4084S,2019MNRAS.489.4090L,2015ApJ...806L..31M,2019A&A...626A.124M,2021A&A...654A..59M,2023MNRAS.521.1408M,2018MNRAS.477.3954P,2023ApJ...948...65L,2001MNRAS.322..231K,2010AJ....139.2679B,2023Natur.613..460L}.}
    \tablecomments{The target type includes UFD: ultra-faint dwarf galaxy; BCG: brightest cluster galaxy; ETG: early-type galaxy; QG: quiescent galaxy; UCMG: ultra-compact massive galaxy; MW: Milky Way. [Z] is the logarithmic mean stellar metallicity relative to the Sun. The applied method for measuring the IMF includes CS: count stars; GCE: galaxy chemical evolution analysis; SPS: stellar population synthesis. The listed mass ranges of the observed stars for the IMF estimation are taken from the corresponding references. The GCE constrain the IMF of stars with a large mass range as explained in Appendix~\ref{Appendix yan2020}. The mass ranges of the stars that contributed to \citet{2024arXiv240411571F} and the SPS spectral-fitting analysis are assumed to be from $0.2~M_\odot$ to $1~M_\odot$ considering the age of the galaxies but should vary for different galaxies and also at different galactocentric distances for resolved observations due to the age and metallicity gradients of galaxies. A precise estimation of the stellar mass ranges of these studies is omitted and is beyond the scope of this paper.
    Assumed shapes of the IMF include SPL: single power law; BPL: broken power law; BT: the bimodal low-mass tapered IMF defined in \citealt{1996ApJS..106..307V}. Specifically, \citet{2020A&A...637A..68Y} assume a power-law IMF that breaks at $0.5~M_\odot$.
    \citet{2018MNRAS.477.3954P} assume a triple power-law IMF with a variable slope between $0.1~M_\odot$ and $0.5~M_\odot$. 
    \citet{2023ApJ...948...65L} assume a fixed slope above $1~M_\odot$ and a variable single slope below $1~M_\odot$.
    \citet{2001MNRAS.322..231K} and \citet{2010AJ....139.2679B} assume a two-part power-law IMF that breaks at $0.5~M_\odot$ with two independent variable slopes.
    $\xi_{\rm MR}\equiv \int_{0.2~M_\odot}^{0.5~M_\odot} \xi(m)\cdot m~{\rm d}m \div \int_{0.2~M_\odot}^{1~M_\odot} \xi(m)\cdot m~{\rm d}m$ is the mass ratio of stars within the ranges of $0.2~M_\odot$ to $0.5~M_\odot$ and $0.2~M_\odot$ to $1~M_\odot$, where $\xi(m)$ is the IMF.
    }
    \tablenotetext{a}{SPS of resolved galaxy regions.}
    \tablenotetext{b}{Analysis of stacked spectra of multiple galaxies or galaxy regions.}
\end{table*}

\begin{figure}\centering
    \includegraphics*[width=\columnwidth]{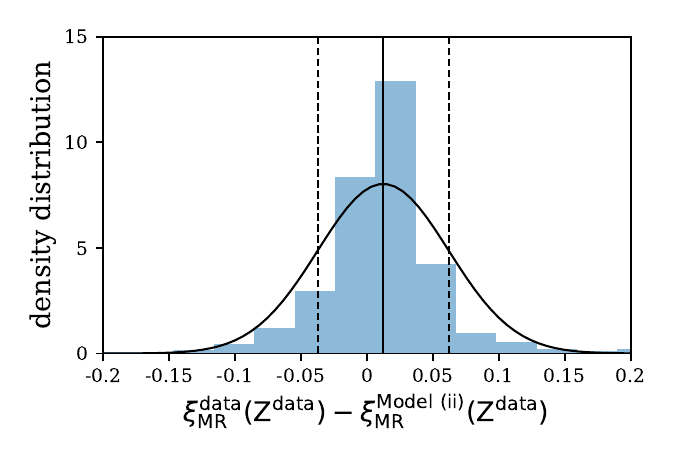}
    \caption{\label{fig: residue_distribution}Distribution of the differences between all the observational estimation of the $\xi_{\rm MR}$ value shown in Figure~\ref{fig: low_mass_IMF} and the expectation of Model (ii) at the same metallicity (histogram). The solid curve is a normal distribution with a mean of 0.01 (solid vertical line) and a standard deviation of 0.05 (dashed vertical lines), the same as the data set.
    }
\end{figure}

A qualitative comparison between Model (i), Model (ii), and the invariant IMF model is given in Section~\ref{sec: Qualitative comparison between models}. Figure~\ref{fig: residue_distribution} shows that the residue between the data and Model (ii) follows a normal distribution with a standard deviation of 0.05, a value smaller than the typical uncertainty of the observations. Systematic differences between the studies and potential observational biases are discussed in Section~\ref{sec: Discrepancies}.

\section{Discussion}\label{sec: Discussion}

\subsection{The metallicity-dependent IMF hypothesis}\label{sec metallicity-dependent IMF}

The development of gas cloud cores sufficiently dense to collapse under self-gravitation to form stars has been known for a long time to depend on the temperature, density, and rate of cooling (metallicity), of the gas (e.g. \citealt{2013pss5.book..115K} and references therein).

A positive correlation between the IMF slope and metallicity is suggested from analytical and numerical simulations (e.g., \citealt{2005ApJ...626..627O,2013ApJ...766..103D,2021MNRAS.508.4175C,2022MNRAS.515.4929G,2023MNRAS.519..688B}). Additionally, a non-linear variation in the IMF with metallicity is proposed owing to the formation of molecules and dust at different metallicities \citep{2020MNRAS.494.2851C,2022MNRAS.509.1959S}. 

Our formulations in equations~(\ref{eq: Yan2020}) and (\ref{eq: Yan2023}) find their origin in observations of Milky Way star clusters \citep{2001MNRAS.322..231K,2002Sci...295...82K}. A critical refinement from \citet{2020A&A...637A..68Y} shifts the focus from the IMF's dependence on log$_{10}$(Z) to a more physically meaningful variation with the stellar metal mass fraction, Z. This adjustment, transitioning from equation~(\ref{eq: Yan2020}) to (\ref{eq: Yan2023}), is particularly relevant when considering a hypothetical metal-free galaxy, where log$_{10}$(Z) becomes negatively infinite. This improvement aligns naturally with observations of metal-rich ETGs, as demonstrated in Fig.~\ref{fig: low_mass_IMF}. 

Theoretical frameworks also posit potential dependencies on other factors (discussed in Section~\ref{sec other physical parameters}). Nevertheless, observationally confirming IMF dependencies on physical parameters like density, pressure, and cloud angular momentum remains challenging compared to the relatively well-recorded influence of metallicity by low-mass stars themselves. With available information and for the purpose of this work, we qualitatively compare different metallicity-dependent IMF models and invariant IMF.

\subsection{Qualitative comparison between models}\label{sec: Qualitative comparison between models}

Figure~\ref{fig: low_mass_IMF} shows that Model (ii) agrees better with the observation. Since a metallicity-dependent model would inherently produce a better fit than an invariant IMF model for having an additional free parameter, we apply the canonical Akaike information criterion (AIC, \citealt{Akaike1998}) to evaluate the relative quality of these models. The criterion identifies the model closest to the ``true model'' in the sense of Kullback–Leibler information and is most effective in predicting new outcomes \citep{Cavanaugh_2019_AIC}.

 Assuming a typical uncertainty of the $\xi_{\rm MR}$ measurements of $\sigma=0.2$ for all the data points in Figure~\ref{fig: low_mass_IMF} and set the residue between measurements and model as $r$, we can calculate the AIC value as ${\rm AIC}=2k-2{\rm ln}(L)$, where $k$ is the number of free parameters of a model and $L=\prod_{i=1}^{i=n}(1-{\rm erf}(|r_i/\sigma|/2^{0.5}))$ is the likelihood function for a model with $n$ being the sample size.

We find that an invariant IMF model with one free parameter has an AIC value of 227. The best-fit parameter is $\xi_{\rm MR}^{\rm best-fit}=0.575$.
Equation~(\ref{eq: Yan2020}) with 2 free parameters has an AIC value of 192. The data-driven best-fit relation is
\begin{equation}\label{eq: M1 best-fit}
\alpha_{1}^{\rm best-fit}\equiv\alpha_{2}^{\rm best-fit}-1=1.62+0.55\cdot{\rm [Z]}.
\end{equation}
Equation~(\ref{eq: Yan2023}), also with 2 free parameters, has an AIC value of 150. The best-fit relation is
\begin{equation}\label{eq: M2 best-fit}
\alpha_{1}^{\rm best-fit}\equiv\alpha_{2}^{\rm best-fit}-1=1.46+54\cdot({\rm Z-Z_{\rm MW}}).
\end{equation}
For stars with ${\rm Z}\approx0$, equation~(\ref{eq: M2 best-fit}) gives $\xi_{\rm MR}({\rm Z})\approx0.46$.
These results suggest that equation~(\ref{eq: Yan2023}), with the smallest AIC value, is far better at representing the truth and making predictions than other models.
Namely, equation~(\ref{eq: Yan2020}), assuming $\Delta\alpha\propto [{\rm Z}]$, is $e^{(150-192)/2}\approx10^{-9}$ times as probable as equation~(\ref{eq: Yan2023}), assuming $\Delta\alpha\propto {\rm Z}$, to minimize the information loss.

The best-fit relations given in equations~(\ref{eq: M1 best-fit}) and (\ref{eq: M2 best-fit}) are based on an inhomogeneous data set shown in Figure~\ref{fig: low_mass_IMF} with a large scatter and different biases. Therefore, it does not necessarily reflect the real variation of the IMF. 
Further tests on the metallicity dependency of the IMF call for a better understanding of observational discrepancies between different studies.

\subsection{Observational discrepancies}\label{sec: Discrepancies}

\subsubsection{Systematic biases}

The observed discrepancies among different studies for the same metallicity could originate from systematic biases introduced by varied assumptions. These include the assumed SFH, the definition of metallicity, the assumed IMF formulation, and other prior assumptions including the stellar evolution model.

For example, the precise stellar mass range contributing to the SPS-based determination of the IMF slope depends on the SFH and stellar metallicity. Galaxies with different metallicities have different colours. Therefore, even for the same underlying IMF and the same population synthesis technique, the IMF measurements could be biased to different stellar mass limits and result in different IMF slopes due to SFH differences. 
% The assumption of a simple stellar population model (where all stars have the same age and element abundances) in SPS studies might introduce bias when applied to stellar populations with diverse SFHs. 
Stellar population models that consider the evolution of star formation density and metallicity enrichment history (cf. \citealt{Haslbauer24}) may offer a resolution to these discrepancies as highlighted in \citet{2021A&A...654A..59M}.

For the definition of metallicity, it is [Z] in most SPS studies but $\rm [Z]+[Fe/H]$ in \citet{2023ApJ...948...65L}. Studies that do not provide a total metallicity but only abundances of individual elements (e.g. \citealt{2023MNRAS.526.4004C,2024arXiv240403939D}) are not included in Figure~\ref{fig: low_mass_IMF}.

In addition, a posterior measure of the shape of the IMF depends on the assumed IMF formulation as a prior (e.g. a power law or lognormal function) while there may be no simple formulation to describe the IMF \citep{1986FCPh...11....1S,2019NatAs...3..482K}. The bimodal low-mass tapered IMF \citep{1996ApJS..106..307V} leads to a bottom-lighter IMF while a single-power-law assumption results in a bottom-heavier IMF (Figures~\ref{fig: IMFs} and \ref{fig: low_mass_IMF}).
Studies from \citet{2018MNRAS.477.3954P} and \citet{2023ApJ...948...65L}, which assume a power-law IMF with variable low-mass end slopes, both yield a steeper $\xi_{\rm MR}$--[Z] relation compared to studies that vary the high-mass IMF instead, probably because a steeper power-law IMF implies more low-mass stars when extrapolating the number of lowest-mass stars according to the assumed IMF (cf. \citealt{2017MNRAS.468..319E}). 
Combining constraints from SPS and gravitational lensing studies, \citet{2020ARA&A..58..577S} suggests that the true IMF for the low-mass stars in massive ETGs might be ``steep but light'', different from the above IMF formulations applied in the literature but in line with a universal mass peak of the IMF at $\approx0.2~M_\odot$ \citep{2018A&A...611A..89L}.

Similarly, star-counting studies of the IMF are model-dependent. The systematically bottom-heavier MW IMF estimated by \citet{2023Natur.613..460L} relative to the canonical \citet{2001MNRAS.322..231K} IMF may be due to a difference of applied stellar mass-luminosity relations \citep{1993MNRAS.262..545K,2023FrASS..1053912X}. 
Star-counting studies for UFDs  \citep{2013ApJ...771...29G,2018ApJ...855...20G,2018ApJ...863...38G,2024arXiv240411571F} adopt fixed prior distributions on the values of interstellar reddening, age, metallicity, and distance of the galaxy, estimated by fitting the observed CMD of a galaxy in previous studies that assume an invariant IMF. However, these are parameters degenerated with the best-fit slope of the IMF and stellar binary fraction.
For example, a higher reddening would suggest a lower binary fraction (cf. \citealt{2023ASPC..534..275O}) and result in a shallower IMF. This could lead to an agreement with Model (ii), strongly favoured by the metal-rich galaxies.
Additional properties that can affect the colour and luminosity of a star but have rarely been accounted for include stellar helium abundance \citep{2010MNRAS.408..999D,2024arXiv240409975J} and velocity distribution of stellar rotation \citep{2022ApJ...938...42H,2024arXiv240108062L}. These lead to additional spreads of the main-sequence stars on the CMD, mistakenly attributed to a higher binary fraction, and therefore, a bottom-heavier IMF.
Follow-up studies need comprehensive uncertainty analysis and a larger range of possible values for the prior distribution of the fitting parameters.

Another potential issue is that the here adopted IMF estimations all neglect mass transfer and binary evolution of stars \citep{2014ApJ...782....7D}. The co-evolution of stars in binary systems can change the colour, spectra, and metal yields of stars, affecting star counting, SPS, and GCE studies, respectively. This needs to be modelled in future studies \citep{2023ascl.soft07035I}.

A comprehensive, homogeneous study considering these systematic effects may help alleviate the discrepancies among different investigations, in which, the stellar mass range over which the IMF slope is determined needs to be carefully examined for individual galaxies with different SFHs.

\subsubsection{Dependencies on other physical parameters}\label{sec other physical parameters}

Alternatively, the disparities among different studies could suggest the influence of additional physical parameters on the shape of the IMF, including the intensity of cosmic rays \citep{2010ApJ...720..226P,2018MNRAS.479.5678F}, gas temperature \citep{1998MNRAS.301..569L,2023ApJS..268...10R}, and consequently, the redshift-dependent cosmic microwave background radiation \citep{2023MNRAS.519..688B}. For example, a redshift dependency of the IMF could lead to a deviation from the prediction of Model (ii) for galaxies with stable or increasing star formation activities (e.g. MW and SMC in Fig.~\ref{fig: low_mass_IMF}).
Star formation density or stellar mass density could also result in a discrepancy \citep{2019MNRAS.489.4090L,2019A&A...626A.124M,2023MNRAS.tmp.3452T,2023Natur.613..460L}, particularly in the low-metallicity regime \citep[see their figure 7]{2021A&A...654A..59M}. 

Considering massive stars, observational evidence suggests that the IMF slope correlates with SFR-related parameters (see references in \citealt{2017A&A...607A.126Y}, \citealt{2018Natur.558..260Z}, and \citealt{2023ApJ...959...88D}). This correlation might extend to low-mass stars as well. 
For example, correlations of the low-mass IMF with SFR and [Mg/Fe] have been observed by \citet{2021A&A...654A..59M}. However, these parameters might not be direct causes of a different shape of the IMF. A higher SFR could bias the IMF slope estimation toward stars that are more massive than $0.5~M_\odot$, which inherently have a steeper IMF slope than lower-mass stars. Similarly, [Mg/Fe] correlates with the formation timescale of a stellar population, and the difference in SFH can affect the estimation of the shape of the IMF. 

Despite efforts, a comprehensive understanding and accurate reproduction of the detailed shape of the IMF and its variation from simulations remains elusive \citep{2024arXiv240407301H}. Challenges include the intricate interplay of feedback mechanisms and cooling, rendering the precise determination of the IMF's characteristic mass difficult \citep{2022MNRAS.515.4929G}. This complexity hinders straightforward comparisons between simulations and observations within specific mass ranges. Studying the IMF across a broad stellar mass range is desirable for comparing general shape variations, but poses challenges due to the difficulty in directly observing both low and high-mass stars within the same stellar population. Potential advancements are expected from future studies in GCE (see Appendix~\ref{Appendix yan2020}).

\subsection{Additional IMF studies}\label{sec: Additional IMF studies}
\subsubsection{The stellar IMF in star clusters}

Compared to studies of the galaxy-wide IMF, the IMF determinations for star clusters encounter additional obstacles due to completeness problems, mass segregation, and dynamical evolution of the star clusters with different ages and orbits within the MW. As a result, many star cluster studies focus on estimating the present-day mass function (PDMF) rather than the IMF (e.g., \citealt{2024arXiv240308850P}).
Correcting for binary fractions, which vary among different clusters, and transitioning from the unresolved-system IMF to the single-star IMF ideally requires detailed N-body simulations of star clusters \citep{2001MNRAS.321..699K}. Despite these challenges, investigations into Galactic young star clusters suggest an invariant IMF for low-mass stars \citep{2021MNRAS.504.2557D}. This universality is likely attributed to the relatively consistent star-forming environment within the closest few kiloparsecs. Notably, the metallicity gradient of the Milky Way is approximately 0.05 [dex/kpc] (\citealt{2005ApJ...618L..95E}, \citealt{2011AJ....142..136L}, \citealt{2015ApJ...808..132H}, and \citealt{2023NatAs...7..951L}). Consequently, the metallicities of the star clusters discussed by \citet{2021MNRAS.504.2557D} likely fall within the range [Fe/H] = $-0.3$ to 0.1. As illustrated in Figure~\ref{fig: Lonoce23} and \ref{fig: low_mass_IMF}, the observed various slopes of the IMF for low-mass stars in this metallicity range are within the margin of uncertainty associated with IMF measurements.

Future investigations of young star clusters could reveal systemic variations in the bottom IMF based on a more extensive dataset and several key improvements. Synthetic photometry models employed for estimating the IMF of a star cluster should incorporate metallicity dependencies. There is also a need to correct the IMF for unresolved binary systems consistently. Young star clusters are likely to exhibit a high binary fraction, which evolves into the lower binary fraction characteristic of the Galactic field \citep{1995MNRAS.277.1491K,1995MNRAS.277.1507K,2015ApJ...800...72T}. This evolutionary process results in an apparently bottom-light IMF for star clusters compared to the average value for the Galactic field stars \citep{2001MNRAS.322..231K}.

For massive metal-poor globular clusters, \citet{2023MNRAS.tmp..617B} determined a single power-law IMF slope of $\alpha=1.65\pm0.3$ for stars between 0.4 and $1~M_\odot$, in agreement with the prediction of equation~(\ref{eq: Yan2023}). Additionally, \citet{2023MNRAS.tmp..617B} did not identify any systematic variation in the IMF slope, aligning with the observations of dwarf galaxies \citep{2018ApJ...855...20G} and the expectation of Model (ii) that the metal dependence of the IMF slope is negligible for metal-poor stars (-2<[Z]<-1, see Figure~\ref{fig: dwarfs}).

The dynamical evolution of globular clusters remains a topic of ongoing debate, with various factors such as initial cluster radii, gas expulsion, and binary formation playing crucial roles but exhibiting considerable uncertainties \citep{2007MNRAS.380.1589B,2012MNRAS.422.2246M,2021MNRAS.504.5778W,2023MNRAS.524.1422F}. Dynamical modelling is necessary to investigate the potential systematic biases in IMF measurements.

\subsubsection{The IMF mismatch parameter}

Many other studies (e.g., \citealt{2012ApJ...760...71C}, compiled in \citealt{2022ApJ...932..103G}) have explored IMF slope variations using SPS. Recently, \citet{2021ApJ...923...43V} proposed a new IMF indicator based on the chromospheric activity of M~dwarfs, yielding results consistent with previous studies based on gravity-sensitive spectral features. Unfortunately, these studies have primarily presented IMF results in terms of the ``IMF mismatch parameter'', a ratio of the expected and observed mass-to-light ratio. This parameter is contingent on stellar remnant mass and, consequently, the IMF slope of massive stars and the poorly constrained SFH of a galaxy \citep{2017A&A...607A.126Y,2019A&A...629A..93Y,2023MNRAS.526.2301D}.

Inherently, the IMF slope of massive stars cannot be constrained in quiescent galaxies lacking massive stars using direct SPS methods, as highlighted in \citet{2019A&A...626A.124M}. However, quiescent galaxies can constrain the IMF of massive stars through GCE calculations. For instance, to reproduce the metallicity and $\alpha$-to-iron element ratio of galaxies with a reasonable SFH, a top-heavy IMF is required for the Galactic bulge and massive ETGs \citep{1990ApJ...365..539M,1994A&A...288...57M,2019A&A...632A.110Y} while a top-light IMF is needed for dwarf galaxies \citep{2020A&A...637A..68Y,2021NatAs...5.1247M}. The variation of the massive end of the IMF alters the stellar-to-remnant mass ratio across different galaxies and even at various radii for the same galaxy. Consequently, constraints derived from the dynamical mass-to-light ratio, coupled with chemical enrichment, play an inseparable role in shaping our understanding of the galaxy-wide IMF of ETGs \citep{2021A&A...655A..19Y,2023MNRAS.526.2301D}. 

\section{Conclusions}
\label{sec: Conclusions}

In this study, we comprehensively discussed how the galaxy-wide IMF for low-mass stars depends on metallicity given the various observations of nearby dwarf galaxies and massive quiescent galaxies. The IMF estimations are based on independent methods, including direct star counts, stellar population synthesis, and galactic chemical evolution. We aimed to reconcile observations assuming different IMF formulations and spanning diverse stellar system masses and ages. We quantified the comparison through the single power-law slope (Section~\ref{sec Compare the IMF slopes}) and the integrated mass ratio between distinct mass ranges (Section~\ref{sec Stellar mass ratio}) as shape variation indicators of the observed IMFs. Our findings highlight the effectiveness of a model incorporating a straightforward linear correlation between the IMF power-law index of low-mass stars and the metal mass fraction [equation~(\ref{eq: Yan2023})], which accurately captures the variation trend of the slope of the galaxy-wide IMF across the majority of observations (Figure~\ref{fig: low_mass_IMF}).

Given that the IMF of massive stars exhibits a strong correlation with the star formation rate or its density, our results suggest a potential for independent variations in the low and high-mass IMF in response to distinct physical parameters, consistent with the proposal by \citet{2018A&A...620A..39J}. Additionally, equation~(\ref{eq: Yan2023}) suggests a relation between the slope of the IMF for stars with different metallicities such that the slope of the IMF for metal-rich stellar populations can be informed by observations of metal-poor galaxies (Section~\ref{sec metallicity-dependent IMF}). We emphasize that for future observational studies of the IMF, it is crucial to explicitly specify the mass range of the observed stars or the stellar mass range their IMF indicator is sensitive to such that different studies can be compared fairly.

Despite potential systematic biases between different studies contributing to observational discrepancies, the formulation of the IMF-variation studied here may be incomplete, and the low-mass IMF is likely influenced by properties of the star-forming environment beyond metallicity considerations (Section~\ref{sec: Discrepancies}). Unfortunately, information regarding other environmental properties, such as star-formation intensity, cosmic-ray density, and temperature, becomes obscured once stars are formed (Section~\ref{sec metallicity-dependent IMF}). The investigation of star formation simulations and observations of star-forming gas clouds could offer valuable insights into the intricate relationship between the IMF and the environment.

%% IMPORTANT! The old "\acknowledgment" command has been depreciated. It was
%% Also note that the acknowledgment environment does not support long amounts of text. If you have a lot of people and institutions to acknowledge, do not use this command. Instead, create a new \section{Acknowledgments}.

\section*{Acknowledgments}
Z.Y.\ acknowledges the support from the Jiangsu Funding Program for Excellent Postdoctoral Talent under grant number 2022ZB54, the National Natural Science Foundation of China under grant numbers 12203021, 12041305, and 12173016, and the Fundamental Research Funds for the Central Universities under grant number 0201/14380049. P.K. thanks the DAAD-East-European-Exchange programme at the University of Bonn and at Charles University for support.

%% in the research for the paper.  Each keyword is checked against the master 

\vspace{5mm}
% \facilities{}


\software{NumPyro \citep{phan2019composable}, NumPy \citep{harris2020array}, SciPy \citep{2020SciPy-NMeth}, Matplotlib \citep{Hunter:2007}.}



\appendix

%These $\alpha$ values represent the galaxy-wide IMF as calculated with the IGIMF theory assuming all stars in the system have one metallicity.
% We obtain posterior distributions for our model parameters $A$ and $\alpha$. After marginalizing $A$, the median value and 15\% (85\%) of the posterior distribution of $\alpha$ were adopted as the IMF slope and its corresponding uncertainty.

\section{Constraints on the IMF of low-mass stars}\label{Appendix yan2020}

In a previous study, \citet{2020A&A...637A..68Y} investigated the chemical evolution of the UFD galaxy Bo\"otes~I, presenting a pioneering effort to constrain the galaxy-wide IMF of low-mass stars using the GCE method. This achievement is made possible by the capability to resolve multiple individual giant stars, allowing the measurement of their element abundances in dwarf galaxies situated within a few hundred kiloparsecs. The observed giant stars, formed at various times and metallicities throughout the galaxy's evolutionary history, serve as a fossil record, providing a detailed chemical evolution history of the galaxy. By comparing this chemical evolution history with models assuming different variations of the galaxy-wide IMF, the study simultaneously constrains the galaxy-wide IMF of high-mass stars that affects the element abundance ratio evolution of the galactic gas reservoir and the galaxy-wide IMF of low-mass stars that affects the metallicity distribution of observed living stars.

Fundamentally, the galaxy-wide IMF of low-mass stars exerts a notable impact on the mean stellar metallicity within the observed galaxy. This influence stems from the fact that giant stars with distinct masses exhibit varying lifetimes. Consequently, these stars must have formed at different epochs, possessing different metallicities to reach their giant phase in the contemporary period. A bottom-heaver galaxy-wide IMF that promotes the formation of lower-mass stars, would lead to an increased observation of metal-poor giant stars formed in earlier stages. Consequently, this scenario would result in a reduction in the mean metallicity of the observed stars.

With the GCE method, the steepness parameter of the IMF-slope--metallicity relation for low-mass stars (e.g. $c_1$ and $c_2$ in equations \ref{eq: Yan2020} and \ref{eq: Yan2023}, respectively) was estimated first in \citet{2020A&A...637A..68Y}. 
Uncertainty was not provided in that and most other GCE studies due to a large number of free parameters that are difficult to analyze, for example, the stellar yields of different elements given by the stellar evolution and explosion models, the star formation efficiency and SFH of a galaxy, the gas accretion, mixing, and outflow (cf. \citealt{2016ApJ...824...82C}). An expensive MCMC calculation is needed for the error estimation (e.g. \citealt{2022ApJ...925...66D,2023MNRAS.526.5084J}) which most studies cannot afford. In addition, the estimated value of the parameter $c_2$ was improved in \citet{2021A&A...655A..19Y} with an updated Solar metallicity (0.02 to 0.0142) and reduced numerical integration error of the GCE code, showcasing the difficulty in estimating the systematic uncertainty for the shape of the IMF. Still, our previous studies inaccurately assume that the stars in the Solar vicinity have the same metallicity as the Sun. Here we consider a mean metallicity of these stars being [Z]$_{\rm MW}=-0.1$ and rescale the steepness parameter from 63 in \citet{2021A&A...655A..19Y} to  
\begin{equation}\label{eq: c2}
c_2=63\times\frac{10^{-2.35}-10^{0}}{10^{-2.35}-10^{-0.1}}\approx79.4,
\end{equation}
where the value $-2.35$ is the iron abundance, [Fe/H], of the target galaxy Bo\"otes~I of the GCE study in \citet{2020A&A...637A..68Y}, and $-0.1$ is the assumed [Z]$_{\rm MW}$ value.

Alternatively, direct star counts offer a means to constrain the galaxy-wide IMF of low-mass stars. However, this approach necessitates the determination of masses for a substantial number of stars and requires correction for observational biases and unresolved binary systems, introducing a considerable level of uncertainty \citep{2013ApJ...771...29G,2017MNRAS.468..319E}. The binary fraction and mass ratio in galaxies with diverse properties may deviate from those in the Solar neighbourhood, posing an additional challenge in this method (cf. \citealt{2011MNRAS.417.1702M}). A joint effort from independent methods including star counting, galaxy spectra synthesis, and chemical evolution modelling will enhance our understanding of the IMF for low-mass stars.


\bibliography{library}{}
\bibliographystyle{aasjournal}



\end{document}